\documentclass{aastex}          

\usepackage{spr-astr-addons}    
\usepackage{url}

\usepackage{natbib}

\begin{document}
%
\title{Relativistic positioning: errors due to uncertainties in the 
satellite world lines}

\shorttitle{<Relativistic positioning errors>}
\shortauthors{N. Puchades and D. S\'aez}

\author{Neus Puchades and Diego S\'aez}
\affil{Departamento de Astronom\'\i a y Astrof\'\i sica, 
Universidad de Valencia, 46100-Burjassot, Valencia, 
Spain.}
\email{diego.saez@uv.es} 


\begin{abstract}
{Global navigation satellite systems use appropriate
satellite constellations to get the 
coordinates of an user --close to Earth-- in an almost inertial 
reference system. We have simulated both GPS and GALILEO constellations.
Uncertainties in the satellite world lines lead to 
dominant positioning errors.
In this paper, a detailed analysis of these errors
is developed inside a great region surrounding Earth.
This analysis is performed in the framework of the so-called 
relativistic positioning systems.
Our study is based on the Jacobian ($J$) 
of the transformation giving 
the emission coordinates in terms of the inertial ones.
Around points of vanishing $J$, positioning errors 
are too large. We show that, for any 4-tuple of satellites,
the points with $J=0$ are located at distances, D, 
from the Earth centre greater than about $2R/3 $, where $R$ is 
the radius of the satellite orbits which are assumed 
to be circumferences. Our results strongly suggest that,
for D-distances greater than $2R/3 $ and smaller than 
$10^{5} \ km $, a rather good positioning 
may be achieved by using appropriate satellite 4-tuples
without $J=0 $ points located in the user vicinity.
The way to find these 4-tuples is discussed for arbitrary 
users with $D<10^{5} \ km $ and, then,
preliminary considerations about satellite navigation 
at $D<10^{5} \ km $ are presented. 
Future work on the subject of space navigation 
--based on appropriate simulations--
is in progress.}

\end{abstract}     


\section{Introduction}
\label{intro}

It is often stated that, 
at distances --from Earth-- greater than 
$D_{max} \sim 2 \times 10^{4} \ km$, positioning 
errors are too big and, consequently, 
satellite navigation based on global navigation satellite systems (GNSS) 
is not feasible
(see \cite{den13} and references cited therein).
This topic about spacecraft navigation is revisited in this paper, where 
the formalism of the so-called 
relativistic positioning systems (RPS) is used. 
In any RPS,  
the location of an user (spacecraft, car on Earth, and so on)
may be achieved by receiving appropriate data from four satellites 
of a certain GNSS (only GPS and GALILEO constellations 
are here considered)

Hereafter, index $A$ labels the four satellites, any other Latin index runs
from $1$ to $3$, and Greek indexes from $1$ to $4$.
Quantities $G$, $M_{\oplus}$, $t$, and $\tau $ stand for the gravitation constant,
the Earth mass, the coordinate time, and the proper time, respectively.
Quantities 
$\eta_{\alpha \beta}$ are the covariant components of the Minkowski 
metric tensor, lengths are given in kilometres and the time unit is defined 
in such a way that the speed of 
light is $c=1$. 

User location requires the choice of a certain reference system. 
It is usually an almost inertial reference. The satellite world lines 
must be known in this reference. Uncertainties in these lines 
lead to positioning errors. The analysis of this type of errors
is the main goal of this paper.

In any RPS,
the user receives codified signals from four satellites at the same time.
After decoding, these signals provide the user with the satellite proper times
at emission. These four proper times are the 
so-called emission coordinates $\tau^{A}$ of the observation event.
From these proper times, the user coordinates in the 
almost inertial reference (hereafter {\em inertial coordinates} $x^{\alpha}$) 
must be calculated; namely, the user position must be found.
Sometimes there are two possible user positions (bifurcation). See
\cite{sch72,abe91,cha94,gra96}. In this case,
additional information is necessary to get the true position 
\citep{col11a,col12a,puc12}.

For photons moving in Minkowski space-time, 
the user inertial coordinates, $x^{\alpha}$, and the
emission ones, $\tau^{A} $, must satisfy the following algebraic 
equations:
\begin{equation}
\eta_{\alpha \beta} [x^{\alpha} - x^{\alpha}_{A}(\tau^{A}) ]  
[ x^{\beta} - x^{\beta}_{A}(\tau^{A}) ] = 0   \ , 
\label{inemis}
\end{equation}    
where $\eta_{\alpha \beta}$ is a diagonal matrix with 
$\eta_{11}=\eta_{22}=\eta_{33}=1 $ and $\eta_{44} =-1 $, 
and the points of the satellite world lines have 
inertial coordinates $x^{\beta}_{A}(\tau^{A}) $, which must be well known functions
of the proper times $\tau^{A}$.
According to Eqs.~(\ref{inemis}), photons follow 
null geodesics from satellite emission to user reception.
These algebraic equations may be solved by using both 
the satellite world lines and the numerical Newton-Raphson 
method \citep{pre99}.  

Eqs.~(\ref{inemis}) may be numerically solved for the unknowns $\tau^{A}$ by assuming that 
the position coordinates $x^{\alpha}$ are known. Thus, the emission coordinates are
obtained from the inertial ones. However, the same equations may be solved to get the
unknowns $x^{\alpha}$ for known emission coordinates $\tau^{A} $. This second 
case gives the inertial coordinates in terms of the emission ones (positioning);
nevertheless, this second numerical solution of Eqs.~(\ref{inemis}) is not necessary since 
there is an analytical formula obtained by \cite{col10a}, which 
gives $x^{\alpha}$ in terms of $\tau^{A}$ for photons moving in Minkowski space-time,
and arbitrary satellite world lines.

We use a manageable approach which leads to
an accurate enough positioning. In this approach, the satellite world lines are appropriate
timelike geodesics of the Schwarzschild space-time, and the photons follow
null geodesics in the Minkowski space-time asymptotic to the Schwarzschild
geometry.

In our approach, there are two types of positioning errors. The first one 
is due to the fact that photons do not move in Minkowski space-time, but 
in the Earth gravitational field.
In the simplest generalization of the above approach,
it may be assumed that both satellites and photons move in the Schwarzschild space-time created 
by an ideal spherically symmetric Earth.
Space-time metrics more general than the Minkowski one
have been considered in 
previous papers (Bahder 2001; \v{C}ade\v{z} \& Kosti\'c 2005; Bini et al. 2008;
Ruggiero \& Tartaglia 2008; 
Teyssandier \& Le Poncin-Lafitte 2008; Delva \& Oplympio 2009; 
\v{C}ade\v{z}, Kosti\'c \& Delva 2010; Bunandar, Caveny \& Matzner 2011;
Delva, Kosti\'c \& \v{C}ade\v{z} 2011).
Metrics including Earth rotation and deviations 
with respect to the spherical symmetry in the Earth mass distribution
might be also considered; nevertheless, since the distance travelled by the photons 
from the satellites 
to any possible user 
is not large, and the Earth gravitational field is weak, the lensing effect 
produced by 
this field --on the photons-- is expected to be small and, 
consequently, positioning errors of the first type should be 
small. A detailed study about these errors
will be presented elsewhere. This paper is devoted to study 
a second type of positioning errors, which 
are due to
uncertainties in the satellite world lines. 
These errors are greater than those due to our assumption 
that photons move in Minkowski space-time (first type).
The equations of the satellite world lines 
are involved in Eqs.~(\ref{inemis}) and, consequently, any 
uncertainty in the satellite motions leads to positioning errors 
when Eqs.~(\ref{inemis}) are solved   
(either analytically or numerically)
to find the user position from the emission coordinates.
 
Any GNSS is based on a certain satellite constellation whose 
world lines (orbits and motions) have been appropriately designed. 
Their equations are known in the almost inertial system of reference.
Hereafter, we say that these satellites form the ideal constellation. 
Of course, these ideal world lines are not followed by the 
true satellites. Even if they are launched to follow them, 
gravity, radiation pressure, and so on, would produce deviations 
with respect to the ideal lines. Sometimes these 
deviations are too large and the satellite world lines 
must be corrected. In this way, 
the space and time deviations of the true world lines 
with respect to the ideal ones would keep 
smaller than a certain upper limit (deviation amplitude), which is hereafter assumed to be 
$10^{-2} \ km$ and $10^{-2} $ time units, respectively.

We simulate the world lines of the GALILEO  and GPS 
background configurations.
The GPS constellation has $n_{s}=24$ satellites 
which move in six 
different orbital planes (four satellites per plane), each plane inclined 
an angle $\alpha_{in}=55 \ deg$
with respect to the equator. 
To obtain around two orbits per day, the satellites are placed at an altitude $h=20200 \ km$. 
We have numerated the satellites in such a way that the satellites 
1 to 4, 5 to 8, 9 to 12, 13 to 16, 17 to 20, and 21 to 24 correspond 
to different consecutive orbital planes. 
The GALILEO constellation is composed by 27 satellites 
($n_{s} = 27$), located in three equally spaced orbital planes 
(9 uniformly distributed satellites 
in each plane). The inclination of these planes is $\alpha_{in}=56 \ deg$  
and the altitude of the circular 
orbits is $h=23222 \ km$; thus, the orbital period is close to 14.2h.
The satellites are numerated as in the GPS case; namely,
satellites 1 to 9, 10 to 18, and 19 to 27 are placed in distinct
consecutive orbital planes.
All the trajectories are assumed to be circumferences whose centres
are located in the Earth centre, which is also the origin of the almost 
inertial reference system used for positioning.

In order to take into account 
the effect of the Earth gravitational field on the satellite clocks,
which run more rapid than clocks at rest on Earth (about 
38.4 microseconds per day for GPS), it is assumed that 
satellites move in Schwarzschild space-time, where 
the circumferences are possible satellite trajectories, 
which are followed with angular velocity
$\Omega = (GM_{\oplus}/R^{3})^{1/2} $. In the asymptotically 
almost inertial system associated to the Schwarzschild
space-time created by Earth, up to first order in the 
small parameter $GM_{\oplus}/R$, the coordinates of a given 
satellite $A$ may be written as follows:
\begin{eqnarray}
x^{1}_{A} &=& R \, [\cos \alpha_{A}(\tau) \cos \psi + \sin \alpha_{A}(\tau) 
\sin \psi \cos \theta] \nonumber \\ 
x^{2}_{A} &=& -R \, [\cos \alpha_{A}(\tau) \sin \psi - 
\sin \alpha_{A}(\tau) \cos \psi \cos \theta] \nonumber \\
x^{3}_{A} &=& -R \sin \alpha_{A}(\tau) \sin \theta \nonumber \\
x^{4}_{A} &=&  \gamma \tau   \ ,
\label{satmot1}
\end{eqnarray} 
where factor $\gamma $ is given by the relation \citep{ash03}
\begin{equation}
\gamma = \frac {dt}{d\tau} =  1 + \frac {3GM_{\oplus}}{2R}  \ ,
\label{ttau} 
\end{equation}  
and the angle
\begin{equation}
\alpha_{A}(\tau) = \alpha_{A0} - \Omega \gamma \tau
\label{satmot2} 
\end{equation}  
localizes the satellite on its trajectory. For any satellite,
angles $\theta $ and $\psi $ and $\alpha_{A0}$ are constant.
The two first angles define an orbital plane in a certain 
GNSS, whereas the third angle 
fixes the position of satellite $A$ at $\tau = x^{4} = 0$.
See \cite{puc12} for more details.

A numerical code with multiple precision has been designed to 
calculate the emission coordinates  
$\tau^{A} $ (unknowns) from the inertial ones (data) by solving  
Eqs.~(\ref{inemis}). It is hereafter referred to as the 
XT-code. This code --based on the Newton-Raphson numerical method--
requires the satellite world line equations; that is to say, 
there must be a subroutine which calculates the inertial coordinates 
of every satellite $x^{\alpha}_{A} $ for any value of $\tau^{A} $.
For the ideal world lines, this calculation
is done by using Eqs.~(\ref{satmot1})--(\ref{satmot2}), but 
the subroutine may work with any known perturbation of these
lines (see below). Other world lines allowed by the 
Schwarzschild geometry, e.g., motions along almost elliptical orbits, 
could be also implemented in the subroutine; 
however, it is not done in this paper since 
our main results about positioning errors would keep
unaltered.

The paper is organized as follows, in Sect.~\ref{sec-2}, an 
analytical formula giving the user coordinates in terms of the 
emission ones, which was derived by \cite{col10a}          
in Minkowski space-time is briefly described. 
Positioning errors due to uncertainties in the 
satellite world lines are studied in Sect.~\ref{sec-3}.
Sect.~\ref{sec-3a} is a theoretical discussion about these errors, 
whereas Sect.~\ref{sec-3b} contains numerical calculations and results.
Finally, a general discussion and 
some comments about perspectives are presented in Sect.~\ref{sec-4}.

\section{Analytical formula for the positioning transformation  $x^{\alpha} = x^{\alpha} (\tau^{A})$}
\label{sec-2}

In this section we use the same compact notation as in 
\cite{col10a}. The inertial coordinates of any event 
are denoted $x \equiv (x^{1},x^{2},x^{3},x^{4}) \equiv (\vec{x},t)$. 

The coordinates ($x^{1},x^{2},x^{3},x^{4}$) of the satellite $A$, at emission time 
$\tau^{A}$, are denoted $\gamma_{A}$. Since the 
world lines of the satellites are known, quantities $\gamma_{A}$ 
may be calculated for arbitrary proper times. The three 
vectors $e_{a} = \gamma_{a} - \gamma_{4} $ define the relative 
positions between satellites $A=a$ and $A=4$. The numeration of the satellites 
and, consequently, the choice of the fourth satellite 
are arbitrary. We may say that vectors $e_{a} $ 
define the internal satellite configuration 
at emission times. There are inertial coordinates characterizing an user 
who receives the times $\tau^{A}$ from the satellites, if and only if, 
the so-called emission-reception conditions, 
\cite{col10a}, are satisfied. These conditions may be written as follows: 
\begin{equation}
\eta_{\alpha \beta} e_{a}^{\alpha} e_{a}^{\beta} > 0, \,\,\,\,     
\eta_{\alpha \beta} (e_{a}^{\alpha}-e_{b}^{\alpha} ) 
(e_{a}^{\beta} - e_{b}^{\beta} ) > 0  \ , 
\label{e_r_c}
\end{equation}  
for any value of indexes $a$ and $b$. 

The general transformation from emission to inertial 
coordinates was derived in Coll, Ferrando \& Morales-Lladosa (2010); it is 
a solution of Eqs.~(\ref{inemis}), which is valid
for arbitrary satellite world lines. In compact formalism, this solution 
may be written as follows:
\begin{equation}
x = \gamma_{4} + y_{\ast} - \frac {y_{\ast}^{2} \chi} 
{(y_{\ast} \cdot \chi)+\hat{\epsilon}\sqrt{(y_{\ast} \cdot \chi)^{2}-
y_{\ast}^{2} \chi^{2}}}   \ ,
\label{emisin}
\end{equation}  
where vectors $\chi $ and $y_{\ast} $ may be calculated from
$e_{1} $, $e_{2} $, and $e_{3} $ (internal satellite 
configuration). The configuration vector $\chi = \ast(e_{1}\wedge 
e_{2}\wedge e_{3}) $ (dual of a double exterior product)
is orthogonal to the hyperplane containing the four $\gamma_{A}$
emission events. Vector $y_{\ast} = (\xi,H)/(\xi \cdot \chi) $,
where $(\xi,H)$ stands for the interior product,
may be calculated from any arbitrary vector $\xi $ satisfying the 
condition $\xi \cdot \chi \neq 0$ and from the bivector
$H = [(e_{a} \cdot e_{a})/2] E^{a} $, where 
$ E^{1} = \ast(e_{2} \wedge e_{3}) $,
$ E^{2} = \ast(e_{3} \wedge e_{1}) $, and  
$ E^{3} = \ast(e_{1} \wedge e_{2}) $. Finally, 
quantity $\hat{\epsilon}$ (orientation of the emission coordinates at $x$) 
can only take on the values $+1$ and $-1$.

In practice, our numerical codes have been designed by using tensor components in the 
almost inertial system of reference; this procedure requires a change of notation. 
From the compact notation used in Eq.~(\ref{emisin}) --which is very appropriate 
for many purposes-- we have passed to index notation (tensor components).
The basic formulae necessary to do this change were explicitly given in a
recent paper by \cite{col12a}. By using index notation, we have built up a numerical code
--based on Eq.~(\ref{emisin})-- which, for given emission coordinates $\tau^{A} $, 
allows us the calculation of the user inertial coordinates $x^{\alpha} $. This 
code is hereafter referred to as the TX-code. Of course, as in the case of
the XT-code described above, 
a subroutine calculating the inertial coordinates 
of every satellite at given values of $\tau^{A} $ is necessary. This
subroutine has the same structure in both codes since it has been
designed to find points of the satellite world lines.

Since the transformation defined by Eq.~(\ref{emisin}) is the 
solution of Eqs.~(\ref{inemis})
for the unknowns $x^{\alpha} $, and Eqs.~(\ref{inemis})
express that 
the distance from $\gamma_{A}$ to $x$ vanishes (in Minkowski space-time), two  
types of solutions may be obtained. The first type corresponds to signals 
{\em emitted from the satellites} at times $\tau^{A}$ and received by an user, at the same time $t$,
at position $\vec{x}$ (emission or past-like solutions). The second type describes 
a signal emitted from position $\vec{x}$ at time $t$ and {\em received 
by the satellites} at times $\tau^{A}$ (reception or future-like solutions). 
Only the first type is significant for positioning. 

In \cite{col10a}, 
it was proved that, for $\chi^{2} \neq 0$, there are two sets 
of inertial coordinates corresponding to $\hat{\epsilon}=+1$ and 
$\hat{\epsilon}=-1$. Moreover, for $\chi^{2} < 0$, 
only one of the two sets of inertial coordinates corresponds to a 
positioning solution. In the case $\chi^{2} > 0$, the number of positioning 
solutions may be either two or zero, in the first case, 
there are two different receptors 
(located at different places),
which would receive the same four emission times from the same 
satellites. In the second case, there are two future-like solutions.
Finally, for $\chi^{2} = 0$ there is only a single positioning 
solution. 

Given four proper times $\tau^{A} $ compatible with 
conditions (\ref{e_r_c}), 
our TX-code calculates all the positioning (past-like) solutions of 
Eqs.~(\ref{inemis}).

\begin{figure*}[tb]
\begin{center}
\resizebox{.4\textwidth}{!}{%
\includegraphics{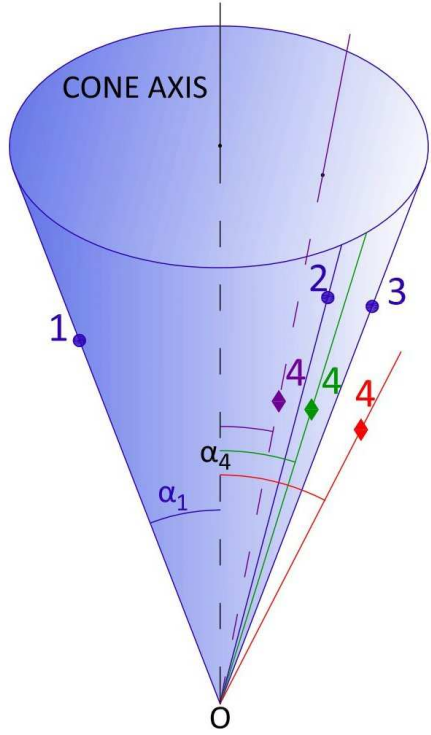}
}
\caption{User-satellites configurations and Jacobian values: 
the cone is generated by the user $O$ and the three satellites 
1, 2, and 3. For the green 
position of satellite 4, the four satellites
are in the surface of the same cone with vertex at $O$, where $J$ vanishes.
Satellite 4 is not on the cone surface for the 
red and fuchsia positions, for which $J$ does not vanish at $O$.}
\label{figu1} 
\end{center}      
\end{figure*}   

\begin{figure*}[tb]
\begin{center}
\resizebox{.5\textwidth}{!}{%
\includegraphics{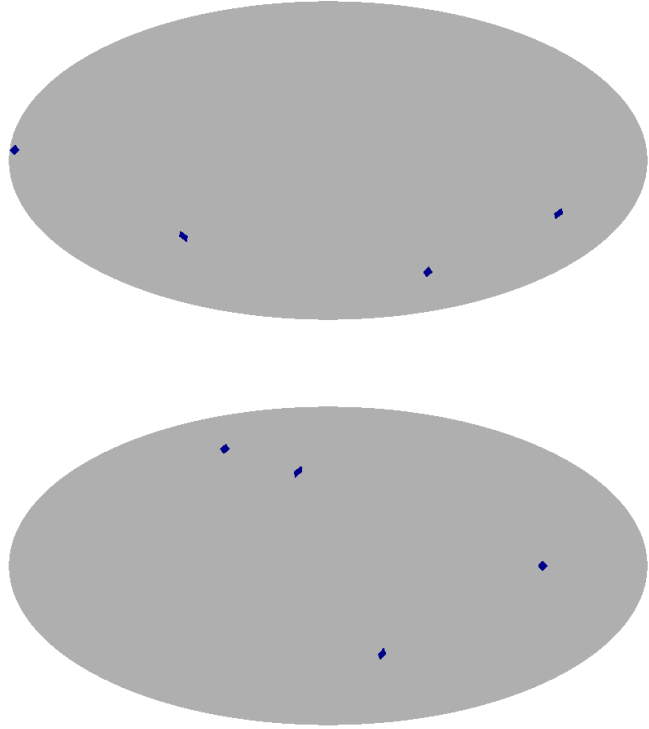}
}
\caption{GALILEO satellites 2, 5, 20, and 23 are observed --from point $E$-- 
in the blue pixels at inertial times $t = 19 \ hours$ (top) and  $t = 25 \ hours$ (bottom)} 
\label{figu2} 
\end{center}      
\end{figure*}

\begin{figure*}[tb]
\begin{center}
\resizebox{1.\textwidth}{!}{%
\includegraphics{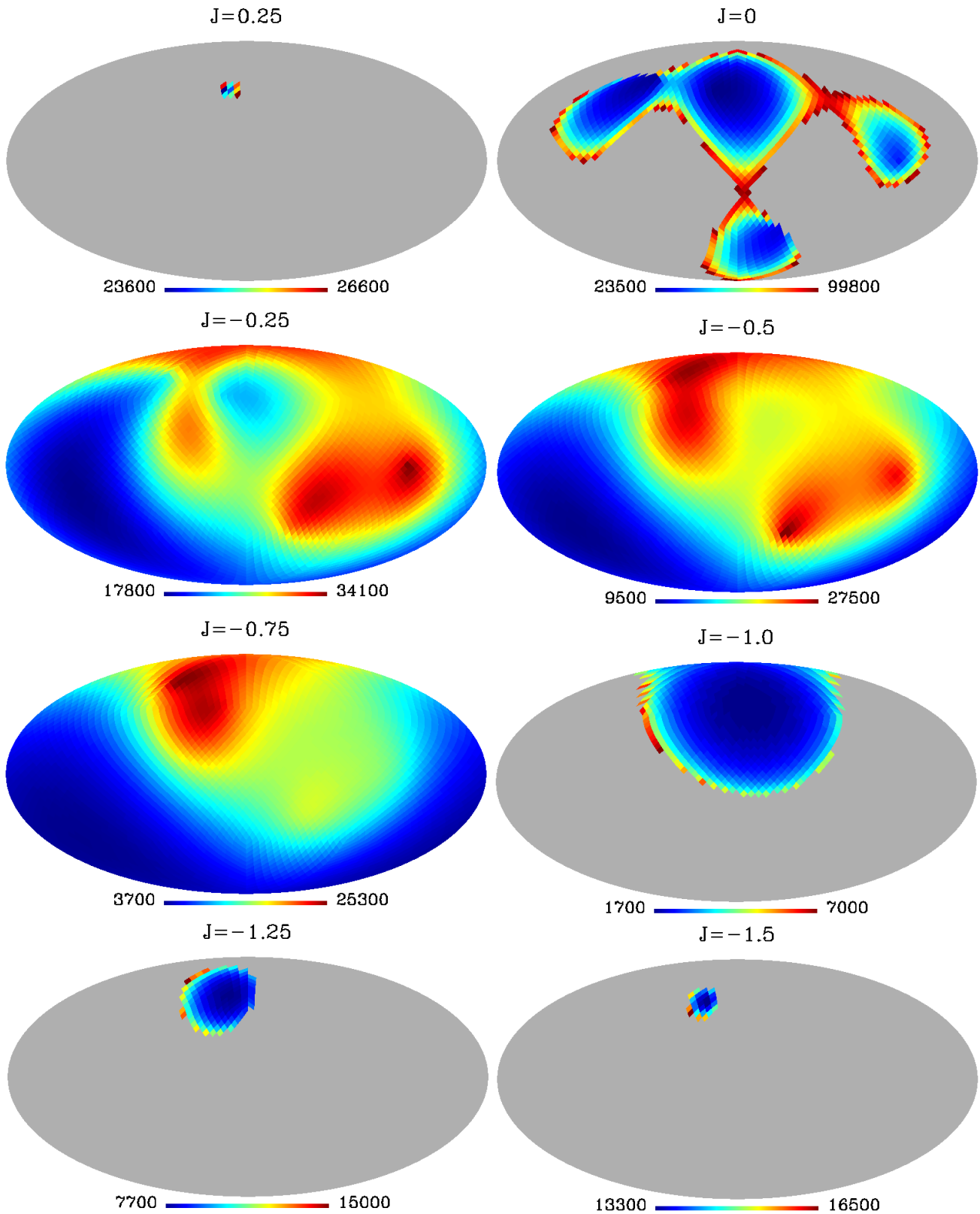}
}
\caption{HEALPIx mollwide maps for the bottom configuration of 
Fig.~\ref{figu1}. The quantity represented is 
the distance, $L_{1} $,  from $E$ to the closest point 
where the Jacobian $J$ takes on the fixed value 
displayed above the map}
\label{figu3} 
\end{center}      
\end{figure*}

\begin{figure*}[tb]
\begin{center}
\resizebox{.5\textwidth}{!}{%
\includegraphics{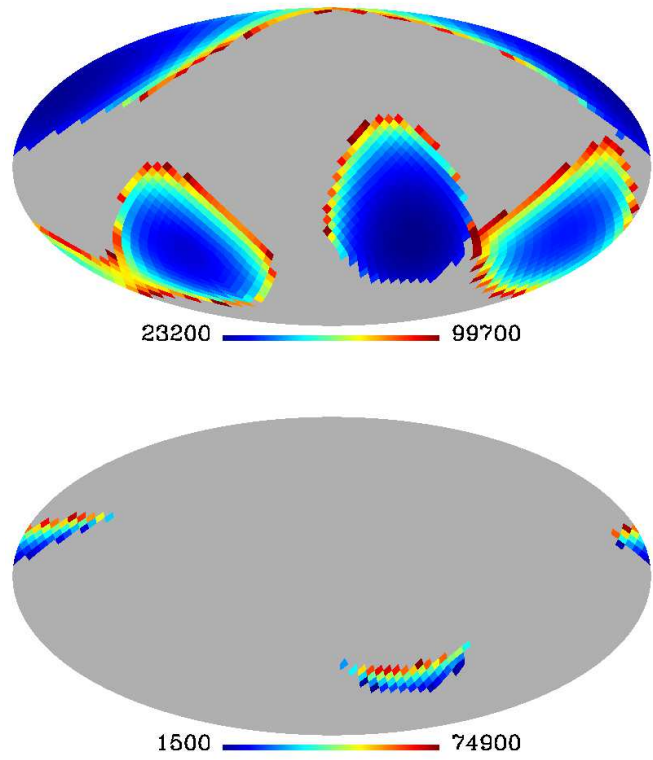}
}
\caption{Top: Same representation as in Fig.~\ref{figu3} for the top
configuration of Fig.~\ref{figu2} and $J=0 $. Bottom: HEALPIx mollwide map
of the distance, $L_{12}$,  from the first to the second point where $J=0 $  
(same configuration as in the top panel).}
\label{figu4} 
\end{center}      
\end{figure*}

\begin{figure*}[tb]
\begin{center}
\resizebox{1.\textwidth}{!}{%
\includegraphics{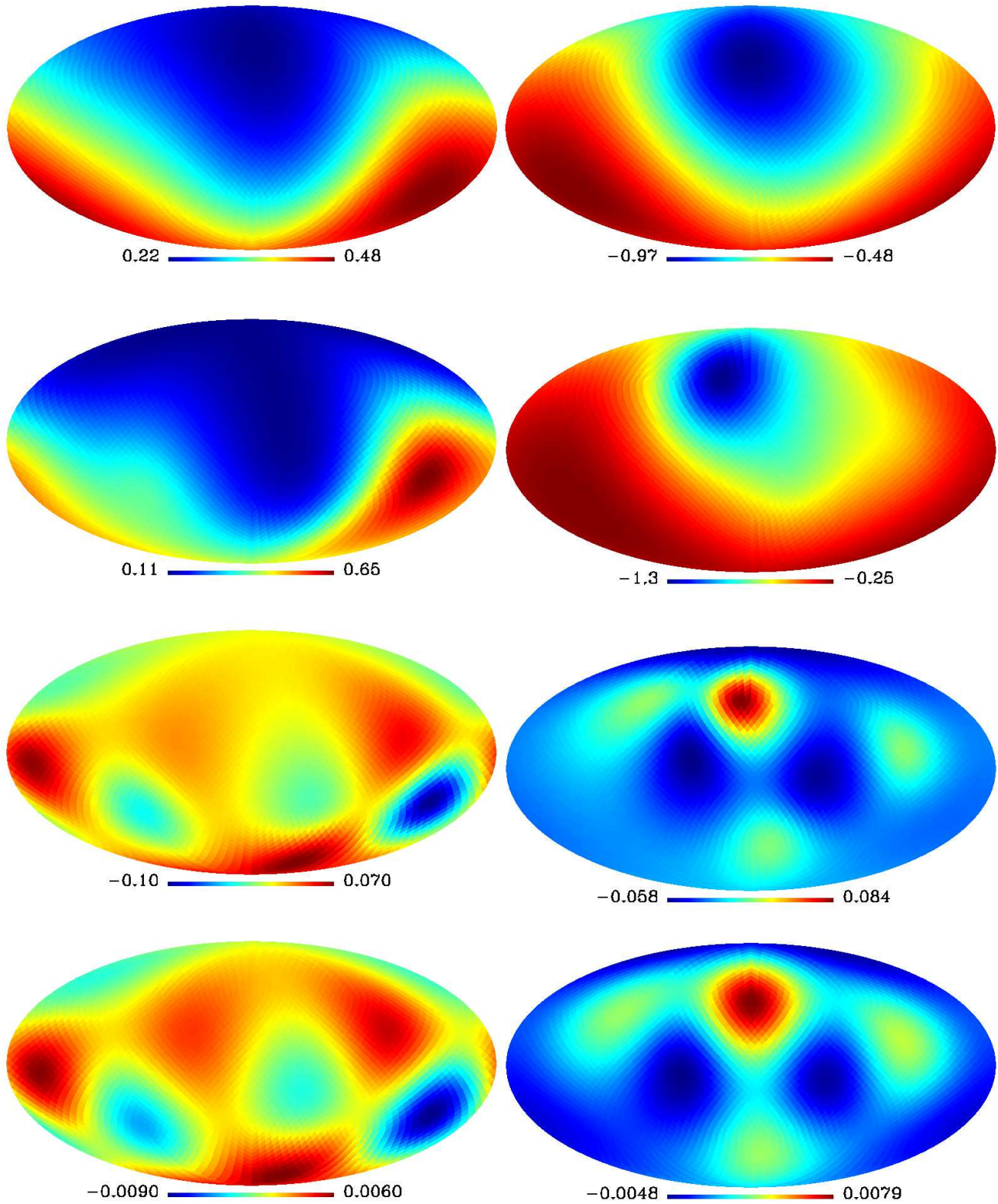}
}
\caption{Left (right): HEALPIx mollwide maps of the $J$ values on 
spheres with different radius for the top (bottom) 
configuration of Fig.~\ref{figu2}. From top to bottom, 
the radius of the spheres in kilometres are 
$6378 = R_{\oplus} $, $1.5 \times10^{4}$, 
$5 \times10^{4}$ and $9 \times10^{4} $.
}
\label{figu5} 
\end{center}      
\end{figure*}   

\begin{figure*}[tb]
\begin{center}
\resizebox{1.\textwidth}{!}{%
\includegraphics{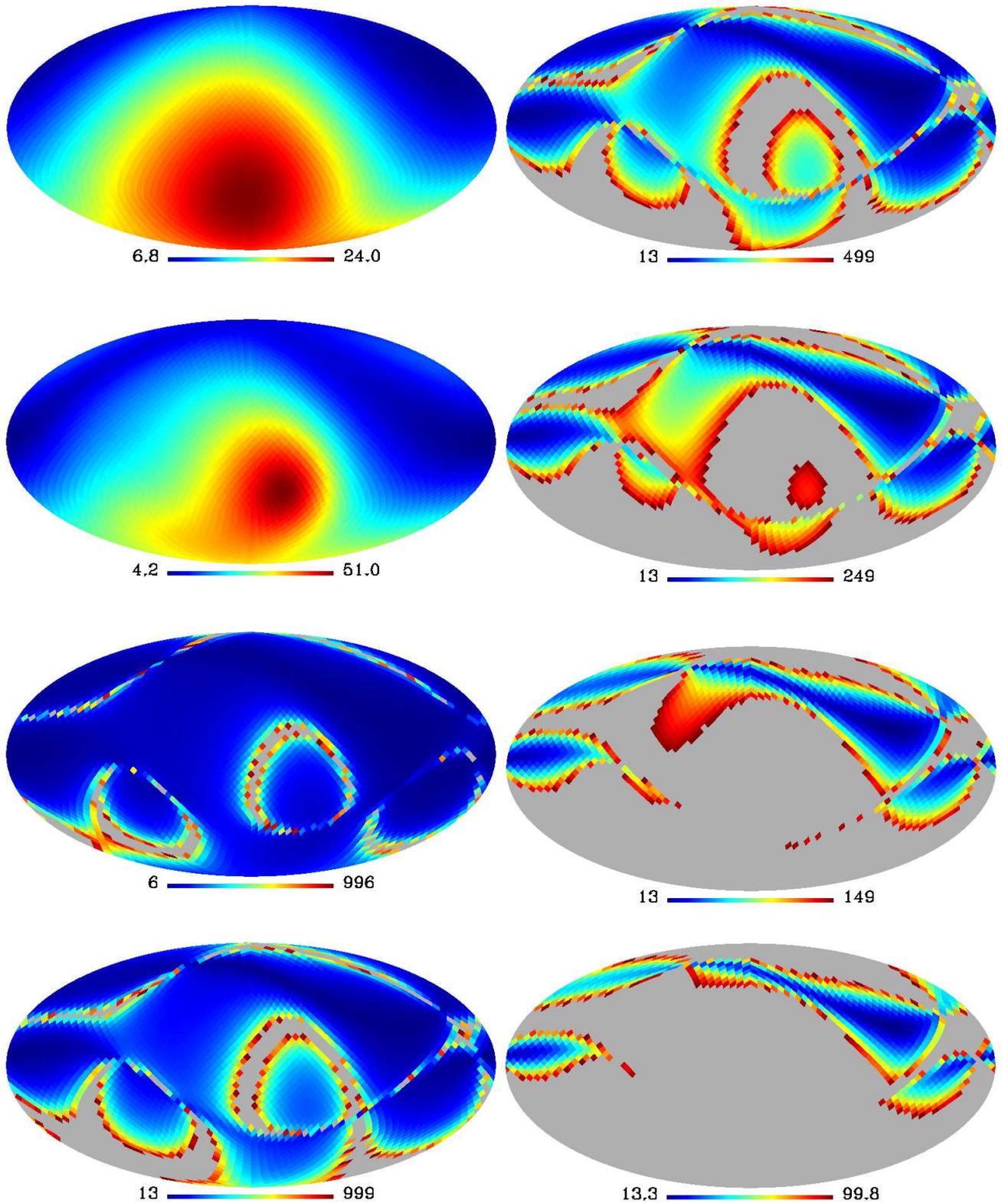}
}
\caption{Left: HEALPIx mollwide maps of $\Delta_{d}$ values, in meters, for
the same spheres and configuration as in the left panels of Fig.~\ref{figu5}. 
Gray (colored) pixels in the middle-bottom and bottom panels are characterized by the 
condition $\Delta_{d} \geq 10^{3} \ m$ ($\Delta_{d} < 10^{3} \ m$). The scale separating 
gray and colored pixels is $10^{3} \ m$. Right: same sphere 
and configuration as in the left bottom panel. From top to bottom, 
the scales, in meters, separating gray from colored pixels are $500$, $250$,
$150 $ and $100 $
}
\label{figu6} 
\end{center}      
\end{figure*}

\begin{figure*}[tb]
\begin{center}
\resizebox{1.\textwidth}{!}{%
\includegraphics{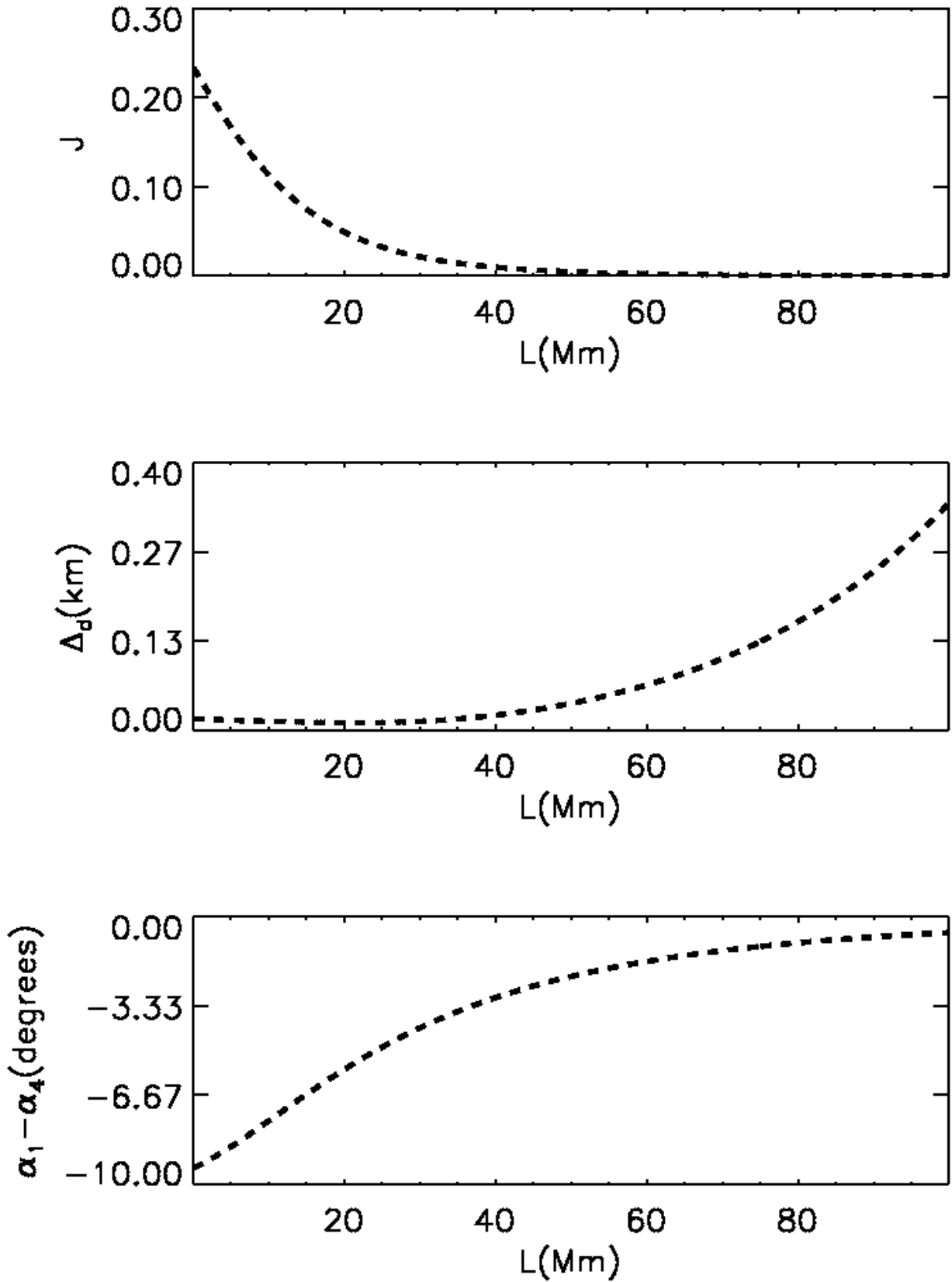}
}
\caption{The values of $J$ (top), $\Delta_{d} $ (middle) and $\alpha_{1} - \alpha_{4} (bottom)$ 
are represented 
along a certain direction from $L=0$ (point $E$) to $L=10^{2} \ Mm$.
The chosen segment, where $J$ does not vanish,
corresponds to the top configuration of 
Fig.~\ref{figu2}. }
\label{figu7} 
\end{center}      
\end{figure*}   

\begin{figure*}[tb]
\begin{center}
\resizebox{.95\textwidth}{!}{%
\includegraphics{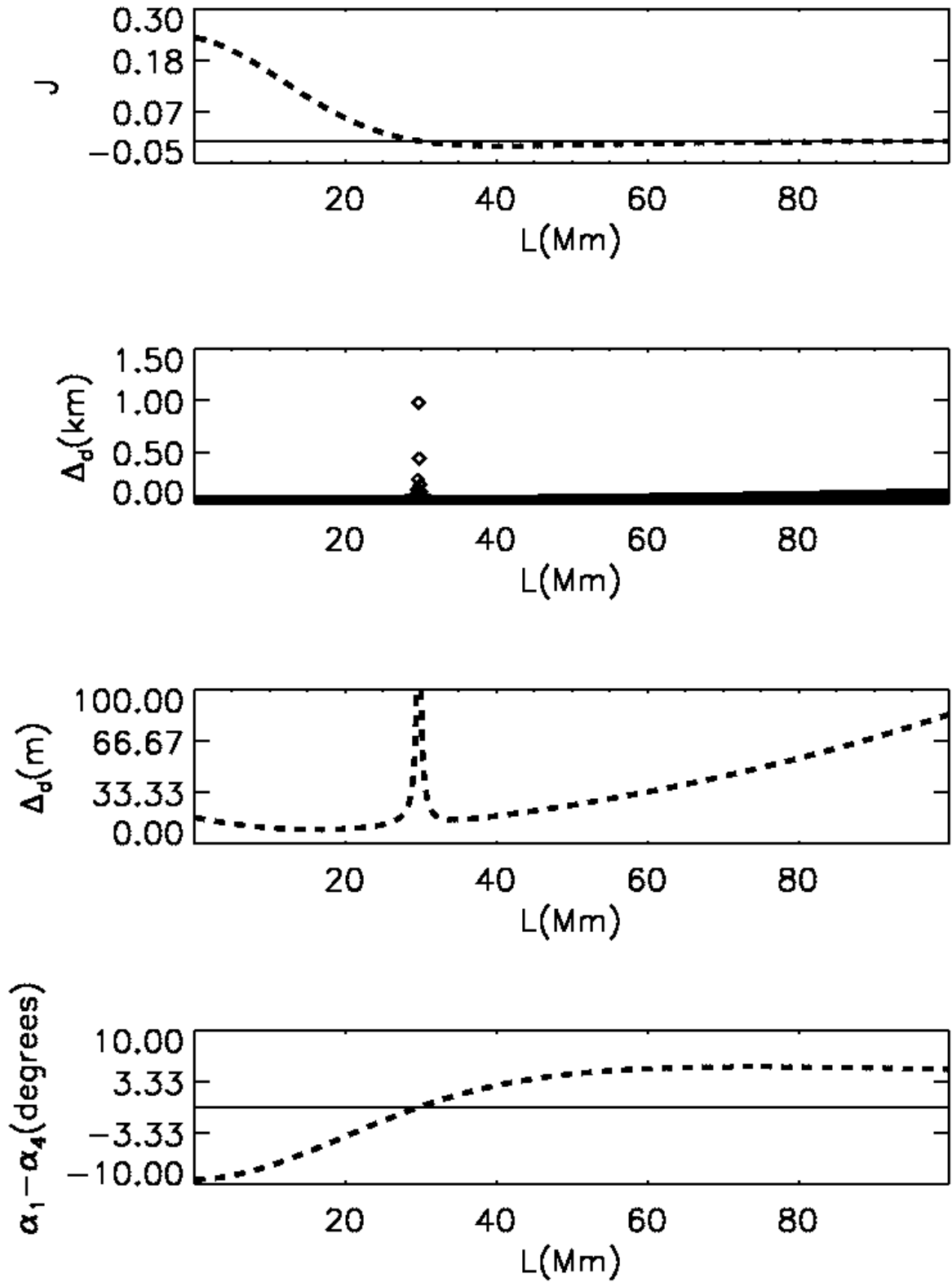}
}
\caption{Same quantities and configuration as in Fig.~\ref{figu7},
for a direction where $J$ vanishes once. In this Figure, the two
middle panels are $\Delta_{d} $ representations. The middle-top panel
includes all the large $\Delta_{d} $ values obtained close to the 
point of vanishing $J$. In the middle-bottom panel, all the $\Delta_{d} $
values greater than $100 \ m$ have been eliminated to make visible 
the positioning errors far from the $J=0$ point.}
\label{figu8} 
\end{center}      
\end{figure*}   

\begin{figure*}[tb]
\begin{center}
\resizebox{1.\textwidth}{!}{%
\includegraphics{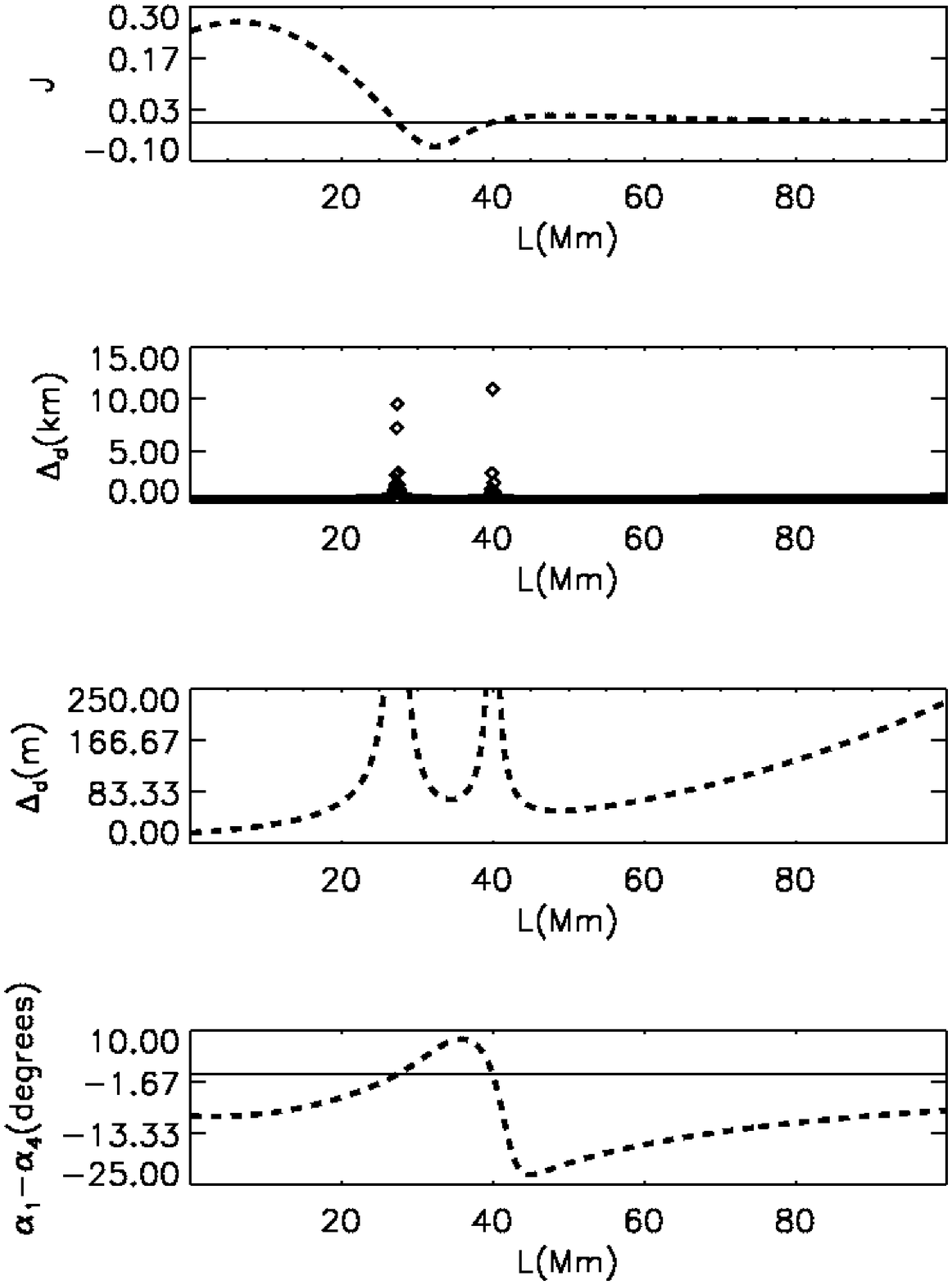}
}
\caption{Same representation as in Fig.~\ref{figu8} for 
a direction where $J$ vanishes twice}
\label{figu9} 
\end{center}      
\end{figure*}

\section{Positioning errors due to uncertainties in the satellite world lines}
\label{sec-3}

Positioning errors have been studied in various papers 
(Langley 1999, Puchades \& S\'aez 2011; S\'aez \& Puchades 2013, 2014).
In \cite{lan99}, GPS positioning errors due to the receiver-satellites geometry 
were studied. It was claimed that these errors strongly depend on the 
volume of the tetrahedron formed by the tips of the 
four user-satellites unit vectors.
The larger the volume, the smaller the positioning errors.
In Sec~\ref{sec-3a}, the tetrahedron criterion is justified 
in the framework of relativistic positioning.
Preliminary work in the line of the present paper was presented 
in some workshops \citep{neu11,sae13,saez14}. 
Here, a more general relativistic four dimensional (4D) study is presented,
this study is extended far from Earth to see the size of 
the region where positioning is accurate enough. Outside this region, 
either pulsar navigation methods or other suitable 
techniques \citep{den13} might be useful.

\subsection{Theoretical considerations}
\label{sec-3a}   

With the essential aim of analyzing positioning errors,
let us assume that: (i) users are located inside a sphere
centred at point $E$, whose radius is $10^{5} \ km$. It 
is hereafter referred to as the $E$-sphere.
The centre $E$ is fully arbitrary. For a given user with coordinates 
$x^{\alpha}$, positioning results do not depend 
on the chosen $E$. The spherical inertial coordinates 
of $E$ have been chosen to be 
$r_{E} = R_{\oplus}$,
$\theta_{E} = 60^{\circ}$, and $\phi_{E} = 30^{\circ}$,
where $R_{\oplus}$ is the Earth radius. Hence, 
point E is on the Earth surface, (ii) 
all the users have the same inertial time coordinate; namely,
they belong to the hypersurface $x^{4} = constant$.
This constant time is arbitrary, but it must be 
a few seconds greater than 
the initial time, $t_{in} $, of the GNSS operation; 
thus, the signals emitted by
the satellites may be received by the users, and 
(iii) opaque objects --as Earth--
intercepting the signal broadcast by the satellites are 
not taken into account, these objects 
may be easily considered after,
without modifying our main conclusions about positioning errors.

Under the simplifying condition (iii),
any user  with 
inertial coordinates 
$x^{\alpha} $, satisfying conditions (i) and (ii),
simultaneously receives codified signals 
--with the emission proper times $\tau^{A} $--
from any set of four satellites of the ideal GNSS constellation. 
Moreover, any neighbouring user 
would receive very similar proper times from the same satellites. 
These facts strongly suggest that, in some open set containing the point $P$ 
with coordinates $x^{\alpha} $, there is 
a function $\tau^{A} = \tau^{A} (x^{\alpha}) $ 
which is continuous and has continuous partial derivatives,
in other words, it is a $C^{1} $ function.
Then, according to the  
inverse function theorem, there is a $C^{1} $ 
function $x^{\alpha} = x^{\alpha} (\tau^{A})$ in some 
open set containing $Q $ (image of P), if and only if, the Jacobian 
$J = |\partial \tau^{A} / \partial x^{\alpha}|$ is a non 
vanishing real number at $P$.
Evidently, any user receiving emission coordinates close to
those of point $Q$ (from the four chosen satellites) 
should be close to the user at $P$. This condition requires
a $C^{1} $ inverse function $x^{\alpha}=x^{\alpha}(\tau^{A})$
with continuous partial derivatives at $Q$ and, consequently,
the Jacobian $J$ must be different from zero 
at $P$. A vanishing $J $ at point $P$ suggests 
strong positioning problems around this point.
Large positioning errors are expected --in the region surrounding $P$--
for any consistent definition of these errors.
This expectation has been numerically 
verified (see below).

The Jacobian of the inverse function $x^{\alpha} = x^{\alpha} (\tau^{A})$
is $J^{\prime} = 1/J = | \partial x^{\alpha} / \partial \tau^{A} |$. If this 
Jacobian is a non vanishing real number at $Q$, $J$ is also a non 
vanishing real number at $P$ and positioning is possible

In order to calculate the partial derivatives involved in $J$,
we may use Eqs.~(\ref{inemis}) with well defined satellite world lines.
From these four equations, one easily finds the following formula
\begin{equation}
\frac {\partial \tau^{A}}  {\partial x^{\alpha}} =  \frac { \zeta  [x^{\alpha} - x^{\alpha}_{A}(\tau^{A})]}
{\eta_{\mu \nu} U^{\mu}_{A}(\tau^{A})
[ x^{\nu} - x^{\nu}_{A}(\tau^{A}) ]}
\ ,
\label{par_tx} 
\end{equation}   
where $\zeta =1$ for $\alpha = 1, 2, 3 $ and $\zeta =-1$ for $\alpha = 4$.
The inertial coordinates $x^{\alpha}_{A}(\tau^{A})$ and the four-velocity 
$U^{\mu}_{A}(\tau^{A}) = d x^{\mu}_{A}/d\tau^{A}$ of satellite $A$
may be easily calculated, at any given proper time $\tau^{A} $,
by means of Eqs.~(\ref{satmot1})--(\ref{satmot2}). Hence,
given an user with inertial coordinates $x^{\alpha} $, our 
XT-code gives the corresponding emission coordinates $\tau^{A}$ and, 
then, the partial derivatives involved in $J$ may be calculated
by using Eqs.~(\ref{par_tx}) and Eqs.~(\ref{satmot1})--(\ref{satmot2}).

The satellite four velocity may be calculated with the formula
$U^{\mu}_{A}(\tau^{A}) = \gamma^{A} (v^{1}_{A}, v^{2}_{A}, v^{3}_{A}, 1)$,
where $v^{i}_{A}$ are the components of the satellite velocity in the almost inertial 
reference, and $\gamma^{A} = [1-(v_{A})^{2}]^{-1/2}$
is the Lorentz factor of satellite $A$. 
Since the satellite 
speeds are much smaller than unity, $U^{\mu}_{A}(\tau^{A}) $
may be approximated by the four-vector
(0,0,0,1) for any $A$. Hence, the following relation is approximately
satisfied: $\eta_{\mu \nu} U^{\mu}_{A}(\tau^{A})
[ x^{\nu} - x^{\nu}_{A}(\tau^{A}) ] \simeq x^{4}_{A}(\tau^{A})-x^{4}
= - D_{A} $, 
where $D_{A} $ is the distance from the user to the position of satellite 
$A$ at emission time. On account of this relation, Eqs.~(\ref{par_tx}) may be 
rewritten as follows:
\begin{equation}
\frac {\partial \tau^{A}}  {\partial x^{\alpha}} \simeq  \frac { \zeta [x^{\alpha}_{A}(\tau^{A})- x^{\alpha}]}
{D_{A}}
\label{parap_tx} 
\end{equation}     
and, consequently, the Jacobian is the value of the following determinant
\begin{equation} \,\,\,\,\,\,            
\left| \begin{array} {cccc}                                                   
\frac{1}{D_{1}}(x^{1}_{1} - x^{1}) & \frac{1}{D_{1}}(x^{2}_{1} - x^{2}) & \frac{1}{D_{1}}(x^{3}_{1} - x^{3}) & 1 \\
                                   &                                    &                                    &   \\ 
\frac{1}{D_{2}}(x^{1}_{2} - x^{1}) & \frac{1}{D_{2}}(x^{2}_{2} - x^{2}) & \frac{1}{D_{2}}(x^{3}_{2} - x^{3}) & 1 \\
                                   &                                    &                                    &   \\ 
\frac{1}{D_{3}}(x^{1}_{3} - x^{1}) & \frac{1}{D_{3}}(x^{2}_{3} - x^{2}) & \frac{1}{D_{3}}(x^{3}_{3} - x^{3}) & 1 \\
                                   &                                    &                                    &   \\ 
\frac{1}{D_{4}}(x^{1}_{4} - x^{1}) & \frac{1}{D_{4}}(x^{2}_{4} - x^{2}) & \frac{1}{D_{4}}(x^{3}_{4} - x^{3}) & 1 \\
\end{array} \right| 
\label{jacobo_tx}
\nonumber 
\end{equation} 

Let us now calculate the volume of the tetrahedron
formed by the tips of the four user-satellites unit vectors. 
Since the system is not relativistic (small velocities), the volume  
may be calculated in any inertial reference system
(almost invariance under Lorentz transformations). If the coordinate  
origin is chosen to be the user position, 
the coordinates of the tetrahedron
vertexes are $\frac{1}{D_{A}}(x^{\alpha}_{A} - x^{\alpha})$ and,
consequently, the absolute value of the above $4 \times 4$ determinant is 
exactly six times the tetrahedron volume $V_{T} $; namely, the relation 
$V_{T} = |J|/6$ is satisfied. Therefore,
it has been proved that, due to the small satellite speeds,
$|J|$ is well approximated by $6 V_{T} $. This fact 
justifies the study carried out by \cite{lan99}
in the framework of GPS, in which,
$V_{T} $ appears to be correlated with positioning 
errors (dilution of precision).

The partial derivatives $\partial x^{\alpha}/ \partial \tau^{A}$ 
involved in the Jacobian $J^{\prime}$
may be also directly computed from Eqs.~(\ref{inemis}), but it is 
not necessary since $J^{\prime}$ may be calculated by using 
the derivatives given by Eq.~(\ref{par_tx}) and the relation 
$J^{\prime}=1/J$.  

As it was proved by \cite{poz06} and \cite{col12a},  
the Jacobian $J$ vanishes if and only if  
the four satellites are on the same cone surface with
the user in the vertex. This is valid for any satellite configuration,
even for a relativistic one with very high velocities. 
In the sketch of Fig.~\ref{figu1}, configurations with $J=0$ 
and $J\neq 0$ may be distinguished. The user located at vertex $O$ and 
the satellites 1, 2, and 3 (at emission times) generate a cone. 
Quantities $\alpha_{1}$ and $\alpha_{4} $ are the angles between the cone axis 
and the lines of sight of
satellite 1 and 4, respectively. 
Hence, the Jacobian $J$ vanishes at $O$ if and only if 
$\alpha_{1} - \alpha_{4} =0$ (green position), whereas it is different from zero 
for the red (fuchsia) position of satellite 4, in which, this satellite is not
on the cone surface but outside (inside).

It is evident that, for standard low velocity satellite systems 
and $\alpha_{1} = \alpha_{4} $, the tips of the four user-satellites unit vectors are 
on the same plane orthogonal to the cone axis and, consequently,
the relation $V_{T} =0 $ is satisfied.  
Moreover, if the user is very far from the satellites, they
are all in a small solid angle and the tetrahedron volume 
$V_{T}$ is expected to be small.
Our numerical estimates are in agreement with these considerations,
since we have found that, 
close to the points $x^{\alpha}$ where $J$ vanishes and $ J^{\prime}$ diverges, small 
uncertainties in the satellite world lines 
lead to large positioning errors, and we have also verified that,
for users $x^{\alpha}$ located far enough from the
satellites, the Jacobian $J$ is small and, accordingly, the positioning errors are 
big.

Let us first suppose that the satellites move, without uncertainties,
according to Eqs.~(\ref{satmot1})--(\ref{satmot2}). These
equations describe the satellite world lines in the case of
a spherically symmetric non rotating Earth, in the absence of external actions. 
In practice, any realistic satellite 
world line deviates with respect to the ideal ones
given by Eqs.~(\ref{satmot1})--(\ref{satmot2}).
If the ideal world lines are parametrized by means of their 
proper times, the equations of these lines 
may be written as follows: $y^{\alpha} = x^{\alpha}_{A}(\tau^{A})$. Then,
the realistic perturbed world lines may be written in the following form
$y^{\alpha} = x^{\alpha}_{A}(\tau^{A})+ \xi^{\alpha}_{A}(\tau^{A})$ in terms
of the same parameters.
Functions $\xi^{\alpha}_{A}(\tau^{A})$ measure the deviations between 
realistic and ideal world lines. These deviations are unavoidable.
  
The ideal world lines of the satellites are those 
of Sect.~\ref{intro}. The trajectories are circumferences 
travelled as it corresponds to the Schwarzschild space-time.
In the absence of deviations with respect to the ideal lines,         
our XT-code gives the emission coordinates $\tau^{A}$ 
corresponding to any set of inertial coordinates $x^{\alpha} $
and, then,
from the resulting emission coordinates, the TX-code 
(based on the analytical solution of Sect.~\ref{sec-2})
allows us to recover the initial inertial ones. 
The number of figures recovered measures the accuracy of 
our XT and TX codes. Since multiple precision is used 
this accuracy is excellent.

Let us now take 
the above emission coordinates $\tau^{A} $, which are not to be varied since they are 
broadcast by the satellites and received by the user without ambiguity, and 
for these coordinates and the 
perturbed world lines $y^{\alpha} = x^{\alpha}_{A}(\tau^{A})+ \xi^{\alpha}_{A}(\tau^{A})$,
the TX code --based on the TX analytical solution-- 
gives new inertial coordinates $x^{\alpha} + \Delta(x^{\alpha})$.      
Coordinates  $x^{\alpha} + \Delta(x^{\alpha})$
are to be compared with the inertial coordinates $x^{\alpha} $ initially assumed.
Quantity $\Delta_{d} = [\Delta^{2}(x^{1})+\Delta^{2}(x^{2})+\Delta^{2}(x^{3})]^{1/2} $
is a good estimator of the positioning errors produced by the
$\xi^{\alpha}_{A}$ uncertainties of the satellite motions.

It is worthwhile to emphasize that user positions $x^{\alpha} $ and
$x^{\alpha} + \Delta(x^{\alpha})$ correspond to the same emission 
coordinates --which are received from the satellites-- but to different world lines. The
ideal world lines lead to position $x^{\alpha} $ and the perturbed ones give 
$x^{\alpha} + \Delta(x^{\alpha})$. We may then say that the user position is 
$x^{\alpha} $ with an error whose amplitude is given by the estimator $\Delta_{d} $,
which must be computed for realistic perturbations of the ideal world lines.

In next section, 
the Jacobian $J$ and the error estimator $\Delta_{d} $ are 
numerically calculated
for appropriate users located inside the $E$-sphere. For each of them, the same deviations 
$\xi^{\alpha}_{A}$ have been used to perturb the ideal satellite world lines.
The three quantities $\xi^{i}_{A}$ have been written in terms of quantity
$\Xi_{A} = [(\xi^{1}_{A})^{2} + (\xi^{2}_{A})^{2} + (\xi^{3}_{A})^{2}]^{1/2} $
and two angles $\Theta $ and $\Phi $ playing the role of spherical coordinates and, then, 
quantities $\Xi_{A} $, $\Theta $, $\Phi $, and $\xi^{4}_{A}$ have been generated --for each satellite--
as random uniformly distributed numbers in the intervals [$0,10^{-2}$] in $km$, [$0,\pi$], [$0,2\pi$], and  
[$0,10^{-2}$] in time units, respectively; in this way, 
the amplitude of the space (time)
deviations has been assumed to be $10^{-2} \ km$ ($10^{-2}$ time units).
These amplitudes were already proposed in Sect.~\ref{intro}.

\subsection{Numerical results}
\label{sec-3b}

Let us analyze relativistic positioning based on satellite 4-tuples
of the GALILEO GNSS. We have verified that the results corresponding to the GPS case 
are analogous.
Since the period of the GALILEO constellation is $\sim 14.2 $ hours, 
our study is restricted to a time interval with this amplitude.
We have chosen two times ($t_{1} =19 \ h$ and $t_{2} =25 \ h$) inside the second orbital period
after the initial operation time $t_{in} = 0$.
A part of the corresponding 3D hypersurfaces --with equation $t=constant $-- 
has been covered by users located  inside 
the $E$-sphere defined in 
Sect. \ref{sec-3a}. 
Calculations have been done for various 4-tuples of 
satellites. In other words,
the Jacobian and the positioning errors have 
been numerically calculated and appropriately represented
for a suitable set of 4-tuples and hypersurfaces. The main results
derived from these numerical calculations are discussed in this section. 

An appropriate method is used to represent some quantities
in the 3D, $t=constant$, space-time sections. Color bars and an appropriate pixelization 
are necessary. In a previous paper \citep{puc12}, the same 
kind of representation was used to display --in some particular cases-- 
the regions where bifurcation (double positioning) takes place.        

In this paper, as in \cite{puc12} and \cite{sae13}, the
HEALPIx ({\it hierarchical equal area isolatitude pixelization
of the sphere}) package \citep{gor99} is used to 
depict appropriate maps. This pixelization was designed to 
construct and analyze maps of the cosmic microwave background.
It is useful to display any scalar quantity depending on the
observation direction (pixel). 
The number of pixels is $12 \times N_{side}^{2}$, where 
the free parameter
$N_{side} $ takes on even natural values. In our maps,
after balancing angular resolution and computational costs,
we have chosen $N_{side} =16$ ($3072$ pixels).
The angular area of any pixel is $\sim 13.43 $ 
squared degrees. This area is close 
to sixty four times the mean angular area of the full moon.
All the HEALPIx pixels have not the same shape, they
are more elongated in the polar zones
[see \cite{gor99}].

Finally, the pixelized sphere is shown by using the mollwide 
projection, in which, the frontal hemisphere is projected on the 
central part of the figure, and the opposite hemisphere is represented
in the lateral parts. The external edges of these parts
represent the same back semi-meridian
(see e.g., any panel of Figs.~\ref{figu2} to~\ref{figu6}).

The knowledge of the HEALPIx and Mollwide 
techniques allows us 
a right intuitive interpretation of the maps.
According to the color bar, any pixel has a color 
which indicates the value of the scalar quantity
displayed in the map, for the direction
associated to the pixel.
For each pixel (direction), a segment starting at the centre $E$
with a length of $10^{5} \ km$ is considered
(a radius of the $E$-sphere) and, then, 
$10^{3} $ points are uniformly distributed along each segment
to cover the sphere under consideration. Each point 
of this coverage has well defined inertial coordinates
by construction and, consequently, the Jacobian $J$ and the 
error $\Delta_{d}$
may be calculated, from these coordinates, following the 
methods described in Sect.~\ref{sec-3a}.
Let us then design various maps and figures to describe the 
4D distribution of the resulting $J$ and  $\Delta_{d}$
values.

\subsubsection{On the 4D distribution of $J$ values. }
\label{ss1} 

In Fig.~\ref{figu2}, we represent the positions of a 4-tuple 
of GALILEO satellites (2,5,20,23) --as they are seen from $E$-- at two different 
inertial times (given in hours). Some maps displayed below correspond 
to these two configurations
(4-tuple plus time). The chosen times define the hypersurfaces 
$t=19$ and $t=25$. Of course, 
more 4-tuples and hypersurfaces of constant time have been considered,
but the main results may be pointed out by using the 
two configurations of Fig.~\ref{figu2}, which are hereafter 
called top ($t=19$) and bottom ($t=25$) configurations
according to their location in the Figure.

For each direction (pixel), we have numerically calculated 
the $J$ values in the $10^{3} $ points previously selected 
(see above).
From the resulting $J$ values, we may easily estimate 
the distance from $E$ to the first point where $J$ takes on a given value.
This distance (hereafter $L_{1} $) is calculated for all the 
HEALPIx directions and represented in a HEALPIx mollwide map.
Of course, the Jacobian could take on the same value in other 
points located at distances from $E$ greater than  $L_{1} $.
In Fig.~\ref{figu3}, there are eight panels (maps) corresponding to
the bottom configuration of Fig.~\ref{figu2}. In each panel,
we show the distance $L_{1} $ in $km$ (color bar) for the $J$ value 
displayed in the top. Grey pixels correspond to segments where there are 
no points with the 
$J$ value under consideration.

In Fig.~\ref{figu3} one easily see that: (i) the values of $|J|$ are
in the interval [0,2), (ii) for $|J| \simeq 1$, there are abundant 
gray pixels and, moreover, for $|J|>1$, the number of these 
pixels increases
as $|J|$ grows; e.g., we see that the colored  
pixels with $|J|=1.5$ (bottom right panel) are very scarce, and
(iii) in the $J=0 $ map there are gray pixels and 
pixels with $L_{1} \geq 23500 $; hence, at distances smaller 
than $23500 \ km$, the Jacobian $J$ does not vanish 
and positioning accuracy is expected to be good enough (see below).
For other configurations, results are similar. Points (i) and (ii)
are always satisfied and point (iii) is always valid up to 
$L_{1} $ distances close to $2 \times 10^{4} \ km $; hence,
it may be stated that the Jacobian does not vanish for 
$L_{1} $ distances smaller than $2R/3 $, where $R$ is the radius of
the satellite orbits. This last statement is valid for both 
GPS and GALILEO satellites.

Fig.~\ref{figu4} corresponds to the top
configuration of Fig.~\ref{figu2}. In the top map,
the distance $L_{1} $ is represented for $J=0$. From this panel it follows that
the inequality
$L_{1} \geq 23200 \ km$ is satisfied. A very similar conclusion
is obtained from the top right panel of Fig.~\ref{figu3}
($L_{1} \geq 23500 \ km$).
Along some directions, there are two or more $J=0$ points.
The number of these points is hereafter denoted $N_{J}$.
The bottom panel of Fig.~\ref{figu4} shows the distance 
$L_{2} - L_{1}$ between the point where $J$ vanishes the 
first time and the next point with $J=0$, which is 
located at a distance $L_{2} $ from $E$. From this panel, 
it follows that the Jacobian $J$ only vanishes two times ($N_{J}=2$)
for a few directions (colored pixed). It vanishes less than 
twice ($N_{J}<2$) for the directions of the gray pixels.
Sometimes, quantity $J$ does not vanish two times  along any
direction ($N_{J}\leq 1$), as it occurs for the bottom configuration of 
Fig.~\ref{figu2}.                                 

From Figs.~\ref{figu3} and~\ref{figu4}, it follows that 
$J$ vanishes at very different distances from $E$. These distances depend 
on direction. Hence, a satellite moving 
inside the $E$-sphere may approach a point with $J=0$, 
where too big positioning errors are unavoidable. Nevertheless, 
any point having $J=0 $ for a certain 4-tuple should  
have $J \neq 0$ for other satellite 4-tuples and, consequently, 
the satellite could be positioned all along its trajectory 
by choosing the most appropriate 4-tuple 
at each moment (see below for more details).

From top to bottom, Fig.~\ref{figu5} shows the values of $J$
on spheres concentric with Earth whose radius, $ R_{s}$, given in kilometres, 
are $R_{\oplus} = 6378 $, 
$1.5 \times 10^{4}$, $5 \times 10^{4}$ and $9 \times 10^{4} $.
Left (right) panels correspond to the top (bottom) 
configuration of Fig.~\ref{figu2}. 
According to previous comments, for the Earth radius (top) and 
for $R_{s}= 1.5 \times 10^{4} \ km$ (middle top), there are no 
$J=0$ points and, consequently, the minimum and maximum $J$ values
displayed in the color bars have the same sign. 
For these two spheres the values of $|J| $ are all greater than 
$0.11 $. In the spheres with radius $R_{s}= 5 \times 10^{4} \ km$
and $R_{s}= 9 \times 10^{4} \ km$, the minimum and maximum
J values of the color bars have opposite signs, which means 
that $J$ vanishes on these surfaces. Moreover, in the left 
middle-bottom and bottom panels, the Jacobian 
is expected to vanish close to the green-yellow zones of 
the maps, which separate blue from red regions. In these
green-yellow transition zones, quantity $|J| $ must be
small and, consequently, large positioning errors are 
expected. In the blue (close to minimum $J$) and red (close to maximum $J$) regions, which are far 
from $J=0 $ points, the $|J| $ values decrease as $R_{s} $
increases; by this reason, the maximum and minimum given in the color bars 
of the bottom panels are smaller than those of the other panels 
corresponding to smaller $R_{s} $ radius; e.g.,
quantity $|J|$ belongs to the interval ($4.8 \times 10^{-3}$, $9\times 10^{-3}$)
in the bottom panels ($R_{s}= 9 \times 10^{4} \ km$) of Fig.~\ref{figu5},
whereas the $|J|$ values range in the interval 
($5.8 \times 10^{-2}$, $10^{-1}$) for $R_{s}= 5 \times 10^{4} \ km$
(middle bottom panels).

In short, there are local decreases of $|J|$ due to local zeros of the Jacobian,
and a gradual decrease --along any direction-- as the distance to $E$ increases
(see the explanation in Sect.~\ref{sec-3a}).
This fact suggests local and gradual growing of the positioning errors 
as the user moves far away from Earth along any direction.

\subsubsection{On the 4D distribution of ($J$,$\Delta_{d}$) pairs}
\label{ss2} 

The $|J|$ values are small in regions close to 
$J=0 $ points, and also in any region located far from both $J=0 $ points and
positioning satellites (located around Earth); hence, in these regions, 
positioning errors are expected to be large.
Since the error estimator $\Delta_{d} $ does not depend 
on $J$ only, but also on other characteristics of the 
user-satellites configuration, the following question arises:
how large are positioning errors for the $J$ values of Fig.~\ref{figu5}? In order to answer 
this question, we have represented the $\Delta_{d} $ values, on the same
spherical surfaces as in Fig.~\ref{figu5} and for the top configuration of
Fig.~\ref{figu2}. Thus,
the $J$ values displayed in the left panels of Fig.~\ref{figu5}
at a certain level (from top to bottom)  
are associated to the $\Delta_{d} $ values represented, at the same level in the left
panels of Fig.~\ref{figu6}. 
Left panels located at the 
same level in Figs.~\ref{figu5} and~\ref{figu6} correspond to spheres with 
the same radius $R_{s}$. The colors of a given pixel in two associated panels 
allow us to estimate a pair ($J$, $\Delta_{d} $). The values of this 
pair of quantities have 
been numerically calculated for 
the chosen direction (pixel).

In the two high levels of Fig.~\ref{figu6} (top and middle top), 
the left panels correspond to spheres with $R_{s} = R_{\oplus}$ (top)
and $R_{s}= 1.5 \times 10^{4} \ km$ (middle top). In these cases, the 
Jacobian $J$ does not vanish (see above) and the positioning 
errors, in meters, are between $4.2 $ and $51.0 $; that is to say, 
these errors are of the same order as the assumed uncertainties
in the satellite positions ($10 \ m$ amplitude). 
The spheres with $R_{s} = 5 \times 10^{4} \ km$ and 
$R_{s} = 9 \times 10^{4} \ km$ are considered in the left panels
of the two low levels of Fig.~\ref{figu6} (middle bottom and bottom). 
In these panels, large positioning errors are expected in the 
zones located close to the $J=0 $ points (green-yellow zones in 
Fig.~\ref{figu5}; see previous comments). According to our expectations, 
very large errors 
have been numerically obtained in these zones.
These errors have been excluded from the maps by performing a 
cutoff at $10^{3} \ m$. To realize this exclusion, the pixels 
having $\Delta_{d} \geq 10^{3} \ m$ 
have been marked with the gray color, whereas the remaining pixels 
have been colored by using the true $\Delta_{d} $ values and 
the color bar. The comparison of the left middle-bottom and bottom panels of 
Figs.~\ref{figu5} and~\ref{figu6} show that, as it was expected, 
the gray 
zones generated with the cutoff
--in the left low level panels of Figs.~\ref{figu6}--
are located in the same places as the green-yellow zones 
of the corresponding panels of Fig.~\ref{figu5}, which are 
in the vicinity of $J=0 $ points.

In the right panels of Fig.~\ref{figu6}, various cutoffs at 
different $\Delta_{d} $ values have been considered. From top to bottom,
the cutoff is performed at the following values in meters:
$500 $, $250$, $150 $, and $100 $. The four resulting maps display
errors on the sphere with $R_{s} = 9 \times 10^{4} \ km$. For the 
same sphere, the 
cutoff at $10^{3} \ m $ has been represented in 
the left bottom panel of the same Figure; hence, five 
cutoffs of the greatest sphere may be found in Fig.~\ref{figu6}.
The cutoff at $10^{3} \ m $ has been already discussed; in 
this case, it may be stated that positioning errors greater 
than $10^{3} \ m$ appears in pixels close to some 
$J=0 $ point. As the cutoff distance decreases, the gray 
zones around the points of vanishing Jacobian 
become widened, and some pixels 
located far from $J=0$ points become gray. 
It is remarkable that, as it is seen in the bottom right
panel, there is a significant number of pixels with 
positioning errors between $13.3 \ m $ and $99.8 \ m$
at a distance of $9 \times 10^{4} \ km$ from the Earth 
centre. This panel strongly suggests that 
the superimposition of various zones with errors smaller than 
$10^{2} \ m $ --corresponding to  
a few satellite configurations different from those of Fig.~\ref{figu2}-- 
could cover the entire sphere for a radius  
of the order of $10^{5} \ km $. This fact strongly suggests that
the region where spacecraft navigation based on GNSS
is feasible might be enlarged beyond spacecraft-Earth distances of $\sim 2 \times 10^{4} \ km $ 
(see Sect.~\ref{intro}), namely, beyond the region 
where there are no $J=0 $ points and the errors have been proved to be small.
Satellite navigation might be possible up to 
distances from Earth as large as $\sim 10^{5} \ km$,
with errors smaller than $\sim 10^{2} \ m$. The main problem is 
the choice of four satellites at each moment during 
spacecraft navigation (see discussion in Sect.~\ref{sec-4}).

\subsubsection{On the ($J$,$\Delta_{d}$) pairs along particular directions}
\label{ss3} 

In addition to the maps presented in Figs.~\ref{figu3}--\ref{figu6},
which contain relevant information on the distribution of  
$J$ and $\Delta_{d} $ values, let us now study the values of
this pair of quantities 
along particular directions corresponding to the 
top configuration of Fig.~\ref{figu2}. Our study 
of these directions has confirmed some previous conclusions 
about ($J, \Delta_{d} $) pairs based 
on the HEALPIX Mollwide maps of Figs.~\ref{figu5}--\ref{figu6} and, 
moreover, this study has provided us with 
additional information 
about the regions surrounding $J=0$ points, and also about
the possibility of finding the best satellite 4-tuple
(minimum positioning error) for a given user. 

Along the first direction, the Jacobian does 
not vanish ($N_{J} = 0$). The $J$ values are displayed in the top panel
of Fig.~\ref{figu7}. These values range from $J=0.23 $ at $E$
to $J=2.6\times 10^{-4} $ for $L= 10^{5} \ km $. 
The fact that the Jacobian does not vanish is 
particularly evident in the bottom panel of Fig.~\ref{figu7},
where quantity $\alpha_{1} - \alpha_{4}$ is clearly different
from zero, which implies $J \neq 0$ (see above). Moreover, 
quantity $\alpha_{1} - \alpha_{4}$, in degrees, varies from 
$-9.4 $ at $E$ to $-0.6 $ at $L= 10^{5} \ km $.
The error estimator $\Delta_{d} $ is represented, in the middle 
panel, as a function of the distance $L$ to $E$. Quantity 
$\Delta_{d} $, in meters, grows from $17.1 $ at $E$ ($L=0$) to 
$338 $ at $L= 10^{5} \ km $. 
From the top and middle panels it follows that $\Delta_{d} $ grows 
as $J$ decreases; nevertheless, $\Delta_{d} $ depends on 
$J$ and also on other quantities defining the 
user-satellites configuration, which means that $\Delta_{d} $
cannot be calculated by using the $J$ value only. For
$L= 10^{5} $, 
an error of $338 \ m $ is associated to the small values 
$J=2.6\times 10^{-4} $ and $\alpha_{1} - \alpha_{4} = -0.6 \ deg $
(first row of Table~\ref{table}).
According to our expectations, it has been verified that the pixel 
associated to this direction 
is colored 
in the top right panel of Fig.~\ref{figu6} (cutoff at $500 \ m$), 
and gray in the middle-top right panel of the same Figure 
(cutoff at $250 \ m$).                  

Fig.~\ref{figu8} shows the results obtained for a second direction,
which contains only a point, $P_{1J} $, where the Jacobian vanishes ($N_{J} = 1$).
The distance $L_{1J}$ --in kilometres-- from this point to $E$ 
satisfies the relation $29700 < L_{1J} < 29800$. Point $P_{1J} $
is visible in all the panels, in the bottom one,
it is the unique point where $\alpha_{1} - \alpha_{4}$ vanishes.
Of course, the Jacobian vanishes at the same point as it can be 
verified in the top panel. In the middle-top panel, quantity $\Delta_{d}$
takes on very large values close to $P_{1J} $. These values
are not important by themselves, since $\Delta_{d}$ diverges
at $J=0 $ and, consequently, the values of $\Delta_{d}$ 
represented in this panel depend on the distances 
between the points selected for calculations and 
point $P_{1J} $. For closer points we would find greater 
values of $\Delta_{d}$. 
The values of $J$, $\Delta_{d}$, and $\alpha_{1} - \alpha_{4}$ at      
point $E$ are independent of the chosen direction. These values have been 
given in previous paragraph (first direction). 
The size of the segment 
centred at $P_{1J} $ where quantity $\Delta_{d}$ 
is greater than $10^{2} \ m$ is $\Delta L_{1J} \simeq 600 \ km$ 
(second row of Table~\ref{table}). 
For this second direction (see Fig.~\ref{figu8}), 
an error of $84 \ m $ is associated to the values 
$J=-1.2\times 10^{-3} $ and $\alpha_{1} - \alpha_{4} = 4.9 \ deg$
for $L= 10^{5} \ km $. All these results have been 
included in the second row of Table~\ref{table}.
It has been verified that the pixel corresponding to this second direction 
is a colored one         
in the bottom right panel of Fig.~\ref{figu6} (cutoff at $100 \ m$). 

Finally, the results obtained for the third direction are shown 
in Fig.~\ref{figu9}. There are two points $P_{1J} $ and $P_{2J} $
where $J$ vanishes ($N_{J} = 2$). The distances $L_{1J} $ and $L_{2J} $ --in kilometres--
from these points to $E$ 
takes on some value inside the intervals 
(27300, 27400) and (39900, 40000), respectively.
From the bottom panel it follows that $\alpha_{1} - \alpha_{4} $ vanishes twice
along this third direction. The two points with vanishing $J$
are also seen in the top panel. The error estimator 
$\Delta_{d}$ is represented in the middle-top and middle-bottom 
panels using the same criteria as
in Fig.~\ref{figu8}.
Moreover, the size of the segments 
centred at $P_{1J} $ and $P_{2J} $ where quantity $\Delta_{d}$ 
is greater than $10^{2} \ m$ are $\Delta L_{1J} \simeq 8700 \ km$ and 
$\Delta L_{2J} \simeq 5000 \ km$, respectively.
For this third direction 
an error of $229 \ m $ corresponds to
$J=2.1\times 10^{-3} $ and $\alpha_{1} - \alpha_{4} = -8.1 \ deg$
at a distance $L= 10^{5} \ km $ from $E$.
All these data have been  summarized in the third row of Table~\ref{table}.  
We have verified that the pixel associated 
to this third direction is colored 
in the middle-top right panel of Fig.~\ref{figu6} (cutoff at $250 \ m$), 
and gray in the middle-bottom right panel of the same Figure 
(cutoff at $150 \ m$).                

\begin{center}
\begin{table*}
\small
\caption{Analysis of positioning errors along three directions \label{table}}
\begin{tabular}{@{}ccccccc@{}}
\tableline
Direction & $N_{J}$ & $\alpha_{1} - \alpha_{4}$ & $J$ & $\Delta_{d}$ & $\Delta L_{1J}$ & $\Delta L_{2J}$ \\
\tableline
  & & & & & & \\
1 & 0 & -0.6 & $2.6\times 10^{-4}$ & 338 & -- & -- \\
2 & 1 & 4.9 & $-1.2\times 10^{-3}$ & 84 & 600 & --  \\
3 & 2 & -8.1 & $ 2.1\times 10^{-3}$  & 229  & 8700 & 5000 \\
  & & & & & & \\
\tableline
\end{tabular}
\tablecomments{Quantities $\alpha_{1} - \alpha_{4}$ [in degrees], $J$, and $\Delta_{d}$ [in meters], 
have been calculated at the boundary of the $E$-sphere for each direction ($L=10^{5} \ km$). 
The widths of the influence areas $\Delta L_{1J}$ and $\Delta L_{2J}$ corresponding 
to a $\Delta_{d}$-level of $10^{2} \ m $ are given in kilometres.
}
\end{table*}
\end{center}

It is known that $\Delta_{d}$ grows as quantities 
$|J|$ and $|\alpha_{1} - \alpha_{4}|$
approach to zero, but this growing only occurs for values
of $|J|$ and $|\alpha_{1} - \alpha_{4}|$ close enough to zero. 
For values of $|J|$ and $|\alpha_{1} - \alpha_{4}|$ which 
are far enough from zero, the error does not grows --in general--
as quantities $|J|$ and $|\alpha_{1} - \alpha_{4}|$
decrease. On account of these comments,  
let us understand the following statements derived from 
Table~\ref{table}: (i) 
the values of
$|J|$ and $|\alpha_{1} - \alpha_{4}|$ corresponding to direction $1$ (first row)
are very small and much smaller
than those corresponding to directions $2$ and $3$ and, 
moreover,
the error of the first direction  ($338 \ m $) appears to be the greatest one,
and (ii) 
for direction $2$, 
quantities $|J|$ and $|\alpha_{1} - \alpha_{4}|$ 
are smaller than those of direction $3$
but the error 
of direction $2$ ($84 \ m $) is not greater than 
that of direction $3$ ($229 \ m $).
All this means that the 4-tuple used to estimate 
the quantities displayed in Table~\ref{table} is not very good 
for the first direction. Other 4-tuples may work better.
Finally, for directions $2$ and $3$, the 4-tuple 
leads to a better positioning for the smaller values of 
$|J|$ and $|\alpha_{1} - \alpha_{4}|$ (compare 
the second and third rows). These results 
must be taken into account to choose the best 4-tuple 
among a set of them (see next section). Of course, this 4-tuple 
must lead to the most accurated positioning.

From the study of many directions corresponding to different 
$t=constant $ hypersurfaces and distinct 4-tuples of GALILEO 
satellites (including the three directions of 
Figs.~\ref{figu7}--\ref{figu9}), 
the following conclusions have been obtained:
(a) points $P_{J} $ where $J$ vanishes are located at different distances
from $E$. These distances depend on the chosen direction.
Each point $P_{J} $ has a region of influence
surrounding it. In this region, the Jacobian is small and positioning errors 
are large. The user must be outside a certain zone --located inside the 
influence region-- 
to have positioning errors below a given level. 
The size of these zones ranges from hundreds to thousands of kilometres 
depending 
on both the chosen $\Delta_{d}$ level and other parameters characterizing the 
users-satellites configurations, and (b) 
far from the $J=0$ points, quantities $|J|$ and $|\alpha_{1} - \alpha_{4}|$
decrease as the distance $L$ to point $E$ increases 
and, consequently, the positioning errors grow. 
This growing and also the decreasing of $|J|$ and $|\alpha_{1} - \alpha_{4}|$ depend on the 
direction. For $L = 10^{5} \ km $, the error $\Delta_{d}$ --far enough from $J=0 $ points--
ranges from about $10^{2} \ m$ to $10^{3} \ m$. 
Conclusions (i) and (ii) are in agreement with
our previous comments about 
Figs.~\ref{figu7}--\ref{figu9}.

\section{Discussion and prospects} 
\label{sec-4}

This paper has been essentially devoted to the study of the 
positioning errors associated to uncertainties in the satellite
dynamics. 
These errors strongly depend
on the Jacobian of the transformation from 
inertial to emission coordinates \citep{neu11,puc12,sae13,saez14}. 

In a given point of Minkowski space-time, with inertial coordinates $x^{\alpha}$,
the Jacobian $J$ and the error $\Delta_{d}$ may be numerically 
calculated by using the methods described in Sect.~\ref{sec-3a}.
Thus, for an appropriate distribution of points covering a region of the 
Minkowski space-time, we can built up maps of both $J$ and $\Delta_{d}$.
Appropriate coverages of a large sphere surrounding the arbitrary 
point $E$ have been defined in Sects.~\ref{sec-3a} and~\ref{sec-3b} and,
then, the resulting distributions of $J$ and $\Delta_{d}$ have been represented 
for some $t=constant$ hypersurfaces. 
In this way, the positions of the 
points where $J$ vanishes have been found for various 4-tuples of GALILEO 
satellites (two of them have been used to design the Figures). 
Other regions, located far from $J=0$ points, where the Jacobian takes on 
small values have been also studied. Taking into account that $J$ tends to zero as 
the distance to $E$ tends to infinity, these regions are far from $E$.

From all the 4-tuples of GALILEO satellites and the $t= constant$ hypersurfaces
we have studied, the following 
conclusions about the distribution of the $J=0$ points
--for distances $L$ smaller than $10^{5} \ km $-- have been found:
(a) for $L$ smaller than $\sim 2\times 10^{4} \ km $,
the Jacobian $J$ does not vanish and positioning is expected to be 
rather accurate. This is in agreement with previous 
statements about the feasibility of space navigation 
by using GPS satellites (see Sect.~\ref{intro}),
(b) for $L$ greater than $\sim 2\times 10^{4} \ km $, 
quantity $J$ does not vanish along many directions, it
vanishes once for a similar number of directions, and only for scarce directions, 
the Jacobian $J$ vanishes twice o more 
times.
Points with vanishing $J$ may be located at any distance $L$ --from $E$-- greater than 
about $ 2\times 10^{4} \ km $,
and (c) In the intervals where $J$ does not vanish, $|J|$ decreases as 
the distance to $E$ increases, reaching very small values 
far enough from $E$.

For a fixed 4-tuple of GALILEO satellites, 
the distribution of $J=0$ points changes with time; hence, we must look for
these points on a large enough number of $t=constant$ hypersurfaces 
covering, at least, a period of the GALILEO satellites. So, a
good description of the 4D distribution of $J=0$ points 
would be obtained.

Let us now suppose a spacecraft launched from point $E$, which travels 
inside the $E$-sphere along a well known world line. This spacecraft
may be seen as an user of the GALILEO GNSS whose world line is known 
and, consequently,
the Jacobian $J$ and the error $\Delta_{d} $ may be calculated 
--with the methods proposed in this paper-- at any point 
of the world line (for each 4-tuple of satellites). In a certain 
position, the user may be inside the influence area of a $J=0$ point for
a certain 4-tuple, but the same position may be far from any point of this type 
($J=0$) for other 4-tuples. This fact strongly suggests that 
the spacecraft might be positioned by choosing the best 
4-tuple in appropriate pieces of the world line; thus, the proximity 
to zero points of $J$ might be avoided along the complete world line.
If this is possible, our numerical results 
(see Sect.~\ref{ss2}) strongly suggest that,
for the best 4-tuples, the order of magnitude for the
positioning errors --inside the $E$-sphere-- might range from
$10 \ m $ (assumed uncertainty of the GALILEO satellites) to $10^{2} \ m$.

If the spacecraft world line is not perfectly known, we could perhaps
study the Jacobian and the positioning errors inside a 4D tube around
the nominal world line. We could look for the 4-tuples avoiding 
$J$ zeros and leading to 
the minimum errors in appropriated pieces of the tube. This information 
might be used as a map to spacecraft navigation close enough to the 
nominal world line. The user (satellite) 
should never go out the tube.
This map should display the most appropriate 
4-tuples for a set of pieces covering the 4D tube. This procedure 
requires simulations to verify the existence of admissible
4-tuples everywhere along the 4D tube; so the trip would be previously planned.
Let us now describe another 
navigation method which seems to be preferable.

Let us finally discuss if an accurate spacecraft (user) position may be found
from information directly obtained by on board devices. 
Of course, there must be devices to detect the codified electromagnetic signals 
broadcast by all the visible GALILEO satellites.
After decoding the signals, the user has the emission coordinates 
corresponding to these satellites. Many 
4-tuples of visible satellites may be then selected to estimate the 
spacecraft position.
The question is: are the positioning errors 
admissible inside the $E$-sphere for a certain 4-tuple? Since 
the user inertial coordinates $x^{\alpha} $ are not known,
the computer on board cannot calculate  the 
Jacobian $J$ and, consequently, we do not know if the 
user is close to $J=0$ points for the chosen 4-tuple; 
in other words, we cannot say anything about the position accuracy 
obtained from the emission coordinates of a given 4-tuple.
Accurate positioning requires additional information 
obtained from the spacecraft, which should carry devices to get 
--with large enough angular resolution--
the lines of sight of the visible GALILEO 
satellites. 
By using this additional information, the quantities $|J|$ (tetrahedron volume) and
$\alpha_{1} - \alpha_{4}$ may be easily calculated for any 4-tuple. 
On account of the resulting values,
some 4-tuples giving too small values of $|J|$ and
$|\alpha_{1} - \alpha_{4}|$ may be discarded; nevertheless,
as it has been concluded in Sect.~\ref{ss3}, 
the preferred 4-tuple (minimum errors) does not correspond to the maximum values of 
$|\alpha_{1} - \alpha_{4}|$ and $|J|$. 
We are currently looking for an operating criterium to 
select the preferred 4-tuple. It seems that 
such a criterion may require 
$|J|$, $|\alpha_{1} - \alpha_{4}|$, and other
quantities characteristic of the user-satellites configuration,
which have not yet been found. After performing an exhaustive 
study about the selection of the best 4-tuple
--which is beyond the scope of this paper--
the proposed navigation method based on emission coordinates 
plus angle measurements must be simulated 
to prove its feasibility.

In order to test the comments about spacecraft positioning given in
the last paragraphs, the location of a  
GPS (GALILEO) spacecraft with the help of the GALILEO (GPS) GNSS is 
being studied in detail.

\acknowledgments{This research has been supported by the Spanish
Ministry of {\em Econom\'{\i}a y Competitividad},
MICINN-FEDER project FIS2012-33582.
We thank B. Coll, J.J. Ferrando, and J.A. Morales-Lladosa for
valuable comments}

\end{document}